\documentclass[preprint,showpacs,preprintnumbers,amsmath,amssymb]{revtex4}

\usepackage{graphicx}% Include figure files
\usepackage{dcolumn}% Align table columns on decimal point
\usepackage{bm}% bold math

\begin{document}

\vspace{5mm}

\newcommand{\goo}{\,\raisebox{-.5ex}{$\stackrel{>}{\scriptstyle\sim}$}\,}
\newcommand{\loo}{\,\raisebox{-.5ex}{$\stackrel{<}{\scriptstyle\sim}$}\,}

\title{Production of spectator hypermatter in relativistic 
heavy-ion collisions.}

\author{A.S.~Botvina$^{1,2}$, K.K.~Gudima$^{1,3}$,
J.~Steinheimer$^{1}$, M.~Bleicher$^{1}$, and I.N.~Mishustin$^{1,4}$}

\affiliation{$^1$Frankfurt Institute for Advanced Studies, J.W. Goethe 
University, D-60438 Frankfurt am Main, Germany} 
\affiliation{$^2$Institute for Nuclear 
Research, Russian Academy of Sciences, 117312 Moscow, Russia} 
\affiliation{$^3$Institute of Applied Physics, Academy of Sciences of Moldova, 
MD-2028 Kishinev, Moldova} 
\affiliation{$^4$Kurchatov Institute, Russian Research Center,
123182 Moscow, Russia}

%
%\def\MOSCOW{Institute for Nuclear Research, Russian Academy of Sciences,
%117312 Moscow, Russia}
%\def\KURCH{Kurchatov Institute, Russian Research Center, 123182 Moscow, 
%Russia}
%\def\FIAS{Frankfurt Institute for Advanced Studies, J.W. Goethe University,
%D-60438 Frankfurt am Main, Germany}
%
%\affiliation{\FIAS}
%\affiliation{\MOSCOW}
%\affiliation{\KURCH}
%
%\author{A.S.~Botvina}       \affiliation{\FIAS}\affiliation{\MOSCOW}
%\author{I.N.~Mishustin}     \affiliation{\FIAS}\affiliation{\KURCH}

\date{\today}

\begin{abstract}

We study the formation of large hyper-fragments in relativistic 
heavy-ion collisions within two transport models, DCM and UrQMD. Our goal 
is to explore a new mechanism for the formation of strange nuclear systems 
via capture of hyperons by relatively cold spectator matter 
produced in semi-peripheral collisions. We investigate basic characteristics 
of the produced hyper-spectators and evaluate the production probabilities 
of multi-strange systems. Advantages of the proposed mechanisms over an 
alternative coalescence mechanism are analysed. We also discuss how such 
systems can be detected 
taking into account the background of free hyperons. This investigation is 
important for the development of new experimental methods for producing 
hyper-nuclei in peripheral relativistic nucleus-nucleus collisions, which 
are now underway at GSI and are planned for the future FAIR and NICA 
facilities. 

\end{abstract}


\maketitle

\section{Introduction}

Strange baryons (hyperons) were discovered in the 1950-s in reactions 
induced by cosmic rays. 
In nuclear reactions at high energies strange particles (baryons and 
mesons) are produced abundantly, and they are strongly involved in 
the reaction dynamics. 
The specifics of hypernuclear physics is that there is no direct
experimental way to study hyperon--nucleon ($YN$) and hyperon--hyperon
($YY$) interactions ($Y=\Lambda,\Sigma,\Xi,\Omega$). 
When hyperons are captured by nuclei, hypernuclei are produced, 
which can live long enough in comparison with nuclear reaction times. 
Therefore, a nucleus may serve as a laboratory offering a unique opportunity 
to study basic properties of hyperons and their interactions.
Double- and multi-strange nuclei are especially interesting, because
they are more suitable for extracting information about the hyperon--hyperon
interaction and strange matter properties.  

The investigation of hypernuclei allows to answer many fundamental 
questions. 
Here we mention only some of them. First of all, studying the structure 
of hypernuclei helps to understand the structure of conventional nuclei 
too \cite{japan}. These studies lead to an extension of the nuclear chart into 
the strangeness sector \cite{cgreiner,greiner}. Second, hypernuclei 
provide a bridge between traditional nuclear physics (dealing with 
protons and neutrons) and hadron physics. Strangeness is an important 
degree of freedom for the construction of QCD motivated models of strong 
interactions \cite{schramm}. And last but not least, strange particles are 
abundantly 
produced in nuclear matter at high densities, which are realized in the 
core of neutron stars \cite{schaffner}. 
The only way to describe realistically these physical conditions is to 
study the hyperon interactions in laboratory, and select theoretical models 
which pass the careful comparison with experimental data. 

%\vspace*{3mm}

Traditionally, information about hypernuclear interactions is obtained 
from spectroscopy of hypernuclei combined with theoretical analyses. 
Production of kaons is often used for triggering hypernuclei, and by 
using kaon beams one can produce double hypernuclei.  
The theoretical studies are mainly concentrated on calculating the 
structure of nearly cold hypernuclei with baryon density around the nuclear 
saturation density, $\rho_0 \approx 0.15$ fm$^{-3}$. In this work 
we consider relativistic nucleus-nucleus collisions leading to 
copious production of hyperons and analyze new opportunities for hypernuclear 
physics realized in this case. 
As seen in experiments in the GeV domain of bombarding energies \cite{Lopez}, 
$\Lambda$-hyperons are produced mainly in the participant zone, however, they 
have a broad rapidity distribution, so that a certain fraction of them can 
even be found in the spectator kinematical region. As shown by theoretical 
calculations using a coalescence model (see, e.g., \cite{gibuu}), 
some of these $\Lambda$-hyperons may be captured by nuclear spectator 
fragments produced in peripheral collisions. Indeed, experiments with 
light-ion beams at the LBL~\cite{Nie76} and JINR~\cite{Avr88} 
have demonstrated that hypernuclei can be formed in such reactions. 
The production of large excited spectator residues is well established in 
relativistic heavy-ion collisions also, see, e.g., Ref.~\cite{aladin95}. 
At a later stage, these excited residues undergo de-excitation via 
evaporation, fission or multifragmentation \cite{smm}. 
Therefore, we expect that the capture of hyperons by spectators may lead to 
the formation of a big lump of excited matter containing a strangeness 
admixture. In the following, these excited spectators will break-up into 
conventional- and hyper-fragments \cite{bot-poch}. 

We note that in central nucleus-nucleus collisions one can also produce 
light hyper-fragments, as has been recently demonstrated in RHIC experiments 
\cite{rhic}.  However, because of the very high excitation energy released in 
the overlapping zone, it will be only possible to produce very light 
hypernuclei ($A \loo 4$) in this way.

\section{Modelling relativistic nucleus-nucleus collisions}
\subsection{The DCM--QGSM approach}

One of the first models designed to describe the dynamics of energetic 
heavy-ion 
collisions was the intra-nuclear cascade model developed in Dubna 
\cite{toneev83}. In the following we refer to it as the Dubna Cascade Model 
(DCM). The DCM is based on the Monte-Carlo solution of a set of the 
Boltzmann-Uehling-Uhlenbeck relativistic kinetic equations with 
the collision terms, including cascade-cascade 
interactions. For particle energies below 1~GeV it is sufficient to 
consider only nucleons, pions and deltas. The model includes a proper 
description of pion and baryon dynamics for particle production and 
absorption processes. 
In the original version the nuclear potential is treated dynamically, i.e., 
for the initial state it is determined using the Thomas-Fermi approximation, 
but later on its depth is changed according to the number of knocked-out 
nucleons. This allows one to account for nuclear binding. 
The Pauli principle is implemented by introducing a Fermi distribution 
of nucleon momenta as well as a Pauli blocking factors for scattered 
nucleons. 

At energies higher than about 10~GeV, the Quark-Gluon String Model (QGSM) 
is used to describe elementary hadron collisions \cite{toneev90,amelin91}. 
This model is based on the 1/N$_c$ expansion of the amplitude for binary 
processes where N$_c$ is the 
number of quark colours. Different terms of the 1/N$_c$ expansion correspond 
to different diagrams which are classified according to their topological 
properties. Every diagram defines how many strings are created  in a 
hadronic collision and which quark-antiquark or quark-diquark pairs form 
these strings. The relative contributions of different diagrams can be 
estimated within Regge theory, and all QGSM parameters for hadron-hadron 
collisions were fixed from the analysis of experimental data. The 
break-up of strings via creation of quark-antiquark and diquark-antidiquark 
pairs is described by the Field-Feynman method \cite{field78}, 
using phenomenological functions for the fragmentation of quarks, antiquarks 
and diquarks into hadrons. The modified non-Markovian relativistic kinetic 
equation, having a structure close to the Boltzmann-Uehling-Uhlenbeck 
kinetic equation, but accounting for the finite formation time of newly 
created hadrons, is used for simulations of relativistic nuclear collisions. 
One should note that QGSM considers the two lowest SU(3) multiplets in 
mesonic, baryonic and antibaryonic sectors, so interactions between almost 
70 hadron species are treated on the same footing. This is a great advantage 
of this approach which is important for the proper evaluation of the hadron 
abundances and characteristics of the excited residual nuclei. 
The above noted two energy extremes were bridged by the QGSM extension 
downward in the beam energy \cite{amelin90}. 

In the course of a nucleus-nucleus collision strange particles are produced 
in both primary and secondary baryon and meson interactions 
(B+B$\rightarrow$BYK, M+B$\rightarrow$YK). 
The produced hyperons can propagate and re-scatter on other particles, 
and, occasionally, they may be located inside the projectile or target 
spectators. These hyperons can be absorbed by the spectators if their 
kinetic energy in the rest frame of the residual nucleus is lower than the 
attractive potential energy, i.e., the hyperon potential. In this case an 
exited residual system with nonzero strangeness will be formed. As known 
from previous studies the $\Lambda$-hyperon potential at the normal nuclear 
density $V_{\Lambda}(\rho_{0})$ is around -30 MeV. In our simulations we 
calculate the local nucleon density $\rho$ at the hyperon's position by 
taking into account only the nucleons in the vicinity of this hyperon, 
within a radius of 2 fm. This local density is then used to calculate the 
effective potential $V_{\Lambda} (\rho )$. Usually this potential is softer 
than the potential in normal nuclear matter, since spectators are quite 
dilute after primary interactions. 
The density dependence of this potential is parameterized following 
Ref.~\cite{Ahmad:1983re}: 
\begin{equation}
	V_{\Lambda}(\rho)= -\alpha \frac{\rho}{\rho_0} 
     \left[1-\beta (\frac{\rho}{\rho_0})^{2/3}\right] ,
\label{binlam}
\end{equation}
where $\alpha=57.5$ MeV, and $\beta =0.522$. 
In our calculations we follow the propagation of each $\Lambda$-hyperon 
during the whole reaction time, up to about 100 fm/c, and the absorption 
criterion is checked regularly. After the absorption some surrounding 
nucleons may 
escape from the spectator, as a result of their interactions with hadrons. 
Therefore, $V_{\Lambda} (\rho )$ can decrease and the absorbed hyperon may 
become free, or, after new interactions, it may be captured again in 
another part of the spectator.  

%\vspace{5mm}

\subsection{The UrQMD approach}
% {\bf  UrQMD description}

For the investigation of spectator hypermatter formation in high energy 
heavy ion collisions we have also employed 
the Ultra-relativistic Quantum Molecular Dynamics model 
(UrQMD v2.3)~\cite{Bleicher:1999xi,Bass:1998ca}. This non-equilibrium 
transport approach constitutes an effective solution of the relativistic 
Boltzmann equation. The underlying degrees of freedom are hadrons, and 
strings that are excited in 
energetic binary collisions. Mean fields can in principle be taken 
into account in this framework too, but for the present investigation we 
have run the model in 
the so called cascade mode without external potentials.

The nucleon's coordinates are initialized according to a Woods-Saxon profile 
in coordinate space and their momenta are assigned randomly according to the 
Fermi distribution in the rest 
frame of the corresponding nucleus. The hadrons are propagated on straight 
lines until the collision criterion is fulfilled: The collision takes place 
if the covariant relative distance $d_{\rm trans}$ between two particles 
gets smaller than the distance $d_0$ corresponding to the total cross 
section $\sigma_{\rm tot}$ (which depends on the energy 
$s$ and the kind of hadrons $h$), 
\begin{equation}
d_{\rm trans}\le d_{0}=\sqrt{\frac{\sigma_{\rm tot}(\sqrt{s},h)}{\pi}}. 
\end{equation}
The reference frame that is used for 
the time ordering of the collisions is the C.M.-system of the nucleus-nucleus 
collision. However, each individual collision process is calculated in the 
rest frame of the binary collision. 

In the UrQMD model 55 baryon and 32 meson species, ground state particles 
and all resonances with masses up to $2.25$ GeV, are implemented with their 
specific properties and interaction cross sections. In addition, full 
particle-antiparticle symmetry is applied. Until now, isospin symmetry 
is assumed and only flavour-SU(3) states are taken into account. The 
elementary cross sections are calculated by detailed balance, or the 
additive quark model, or are fitted and parametrized according to the 
available experimental data. For resonance excitations and decays the 
Breit-Wigner formalism is employed assuming vacuum parameters.

Towards higher energies, the treatment of sub-hadronic degrees of freedom 
is of major importance. In the present model, these degrees of freedom 
enter via the introduction of a formation time for hadrons produced in 
the fragmentation of strings 
\cite{Andersson:1986gw,NilssonAlmqvist:1986rx,Sjostrand:1993yb}. 
String formation and fragmentation is treated according to the Lund 
model. For hard collisions with large momentum transfer ($> 1.5$ GeV/c) the 
Pythia model is used for the simulation of final states. 
The UrQMD 
transport model is successful in describing the yields and the $p_{t}$ 
spectra of various secondary particles in pp and pA collisions 
\cite{Bratkovskaya:2004kv}. A compilation of results of the recent version 
UrQMD-2.3 compared to experimental data can be found in 
\cite{Petersen:2008kb}.

For our present purposes we have applied the following procedure:
During the UrQMD calculational run, we check for all $\Lambda$s and 
$\Xi$s in time steps of $\Delta t = 0.5 \ \rm{fm}/c$, first selecting 
only particles which are in the rapidity interval  $y_b \pm \Delta y$, 
where $y_b$ is the beam rapidity and $\Delta y=0.267$, the rapidity spread 
associated with the Fermi momenta of nucleons in the initial nucleus.
For such particles the net baryon density in the local rest frame of the 
strange particle is then estimated. This is done by calculating the zero 
component of the net baryon current of all baryons within the same 
rapidity interval. For this purpose we assume that each baryon is represented 
by a Gaussian wave package of width $\sigma = 1$ fm which is Lorentz 
contracted in the direction of motion. 

For the hyperon absorption by spectators the same potential criterion as 
in DCM model was adopted: We calculate the kinetic energy of the strange 
particles, both $\Lambda$ and $\Xi$-hyperons, 
with respect to the rest frame of the projectile nucleus. Whenever this 
energy is less than the corresponding potential, calculated with 
Eq.~(\ref{binlam}), the particle is absorbed and removed from the 
cascade simulation. We have checked that the results do only weakly 
depend on the depth of the hyperon potential, when it is varied in a 
reasonable interval of $\pm 20 \%$. 

%%%%%%%%%%%%%%%%%%%%%%%%%%%%%%%%%%%%%%%%%%%%%%%%%%%%%%%%%%%%%%%%

\section{Formation of hyperon-rich spectator matter}

%\subsection*{\bf 4.1\hspace*{2ex} Previous work}

A detailed picture of peripheral relativistic heavy-ion collisions has 
been established in many experimental and theoretical studies. 
Nucleons from the overlapping parts of the projectile and target 
(participant zone) interact intensively between themselves 
and with other hadrons produced in primary and secondary collisions. 
Nucleons from the non-overlapping parts do not interact strongly, and 
they form the residual nuclear systems, 
which we call spectators. In all transport models 
the production of hyperons is associated with nucleon-nucleon collisions, 
e.g.,  p+n$\rightarrow$n+$\Lambda$+K$^{+}$, or collisions of secondary 
mesons with nucleons, e.g., $\pi^{+}$+n$\rightarrow \Lambda$+K$^{+}$. 
Strange particles are mainly produced in the participant zone, however, 
they can re-scatter and populate the whole momentum space around the 
colliding nuclei. 

In the present paper we are interested mostly in 
hyperons which propagate with velocities close to the initial 
velocities of the nuclei, i.e., in the vicinity of nuclear spectators. 
We assume that such hyperons can be absorbed by 
the spectators if their kinetic energy (in the rest frame of the 
spectator) is lower than the potential generated by neighbouring spectator 
nucleons. Due to the secondary interactions these spectators are 
excited and have a dilute (subnuclear) density. 
As was previously investigated in detail \cite{aladin95,aladin97,EOS}, 
the further evolution of the spectators depends on their excitation 
energy. At lower excitation energies (below $2 \div 3$ MeV per nucleon) 
they de-excite via evaporation and/or fission-like processes, while 
at higher excitation energies they undergo multifragmentation \cite{smm}. 
We expect the same behaviour for the residual nuclei containing hyperons. 

As was also found in previous studies of spectator fragmentation 
\cite{aladin97}, there exists a strong correlation between the excitation 
energy transferred to the spectator matter and the number of high energy 
particles produced in the participant zone. On the other hand, the 
multiplicity of strange particles is also proportional to the number of 
participants. Therefore, we expect that the hyper-spectators will be 
very excited, and finally they will break-up into many fragments, some 
of them will contain the captured hyperons. This process was already 
analysed in Ref.~\cite{bot-poch}. 

\begin{figure}[tbh]
\includegraphics[width=0.8\textwidth]{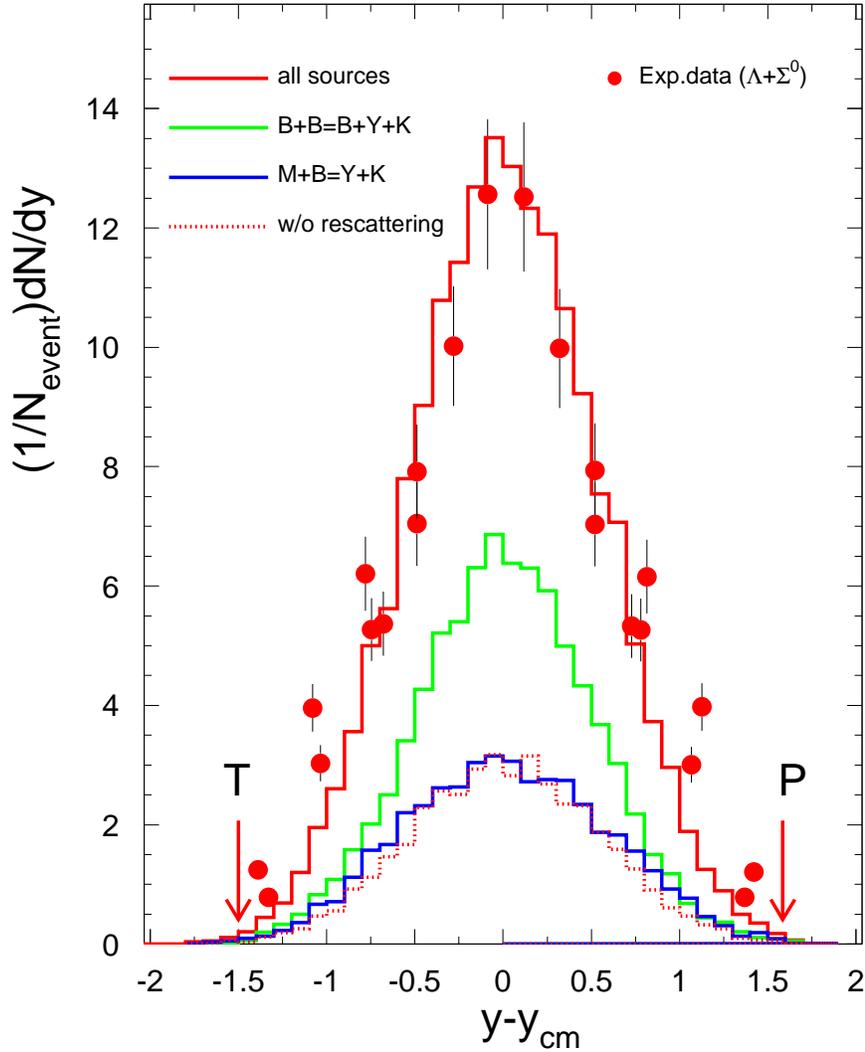}
\caption{\small{
Rapidity distribution of $\Lambda$ and $\Sigma^{0}$ hyperons produced 
in collisions of Au projectile with momentum 11 GeV/c per 
nucleon on Au target. Red circles are experimental data \cite{E896}. 
Histograms show contributions of different processes as predicted by 
DCM. Rapidities of the projectile (P) and target (T) are shown by arrows. 
}}
\label{fig1}
\end{figure}

The DCM model gives a very good description of yields and spectra of 
free hyperons and kaons produced in relativistic heavy-ion collisions. 
For example, in Fig.~\ref{fig1} the calculated inclusive rapidity 
distributions of $\Lambda$ and $\Sigma^{0}$-hyperons are compared with 
experimental data of E896 experiment at AGS \cite{E896}. 
As one can clearly see, the primary nucleon--nucleon interactions, 
which are noted as 'without rescattering', are responsible only for 
a small fraction of the strange particles. 
The secondary interactions of baryons and mesons 
provide the dominant contribution to the production of 
hyperons at all rapidities, 
and they are crucially important for description of the data. 
Similarly, the UrQMD model proved to be very successful in description of 
the strangeness production in high-energy nuclear collisions (see, e.g., 
\cite{Petersen:2008kb}). 

%\vspace{5mm}

\begin{figure}[tbh]
\includegraphics[width=0.8\textwidth]{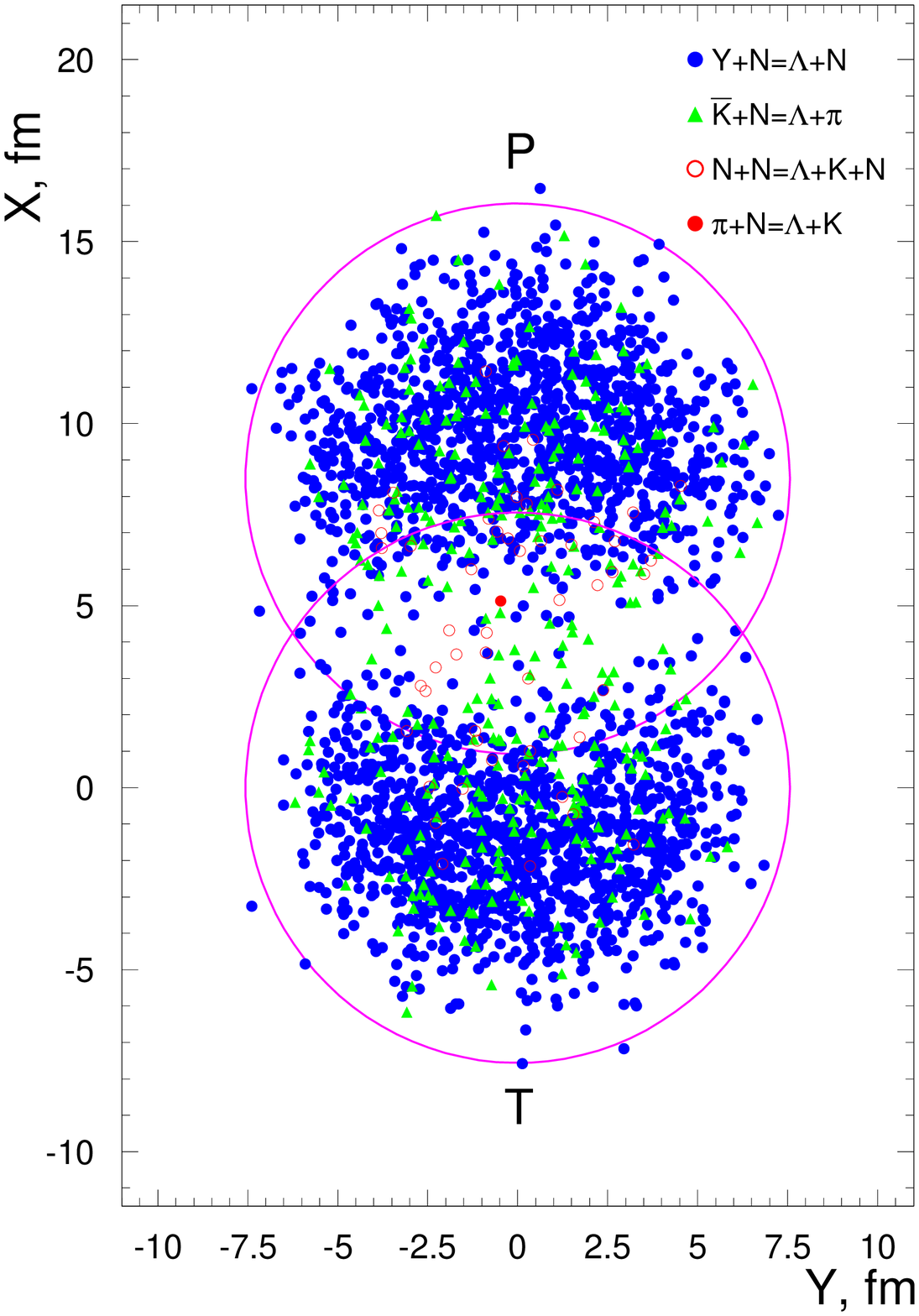}
\caption{\small{
Coordinates of $\Lambda$ absorption points projected into the 
transverse plane perpendicular to the beam axis. 
Results of DCM calculations are shown for Au (20 A GeV) + Au 
collisions with an impact parameter of 8.5 fm. The circles are outer 
contours of projectile (P) and target (T) nuclei at this impact parameter 
Processes creating the absorbed hyperon, such as interactions of 
secondary hyperons, antikaons, nucleons and pions with nucleons, 
are indicated by different symbols. 
}}
\label{fig2}
\end{figure}

Since both the DCM and UrQMD can follow the evolution of all interacting 
particles in space and time, we can predict where and when the 
produced $\Lambda$ is absorbed in a projectile or target spectator. 
As a first example, in Fig.~\ref{fig2} we demonstrate the spatial 
distribution of $\Lambda$-hyperons, which satisfy the capture criterion 
described above. This figure shows the coordinates of absorption 
points projected into the transverse plane (X and Y coordinates), 
perpendicular to the beam axis (Z coordinate). 
The results are presented for Au + Au collisions at a beam energy of 
20 GeV per nucleon and an impact parameter of 
8.5 fm. Such collisions will be studied in future experiments at FAIR 
(Darmstadt, Germany) and at NICA (JINR, Dubna, Russia). The figure 
accumulates the simulation results after 2$\cdot$10$^5$ collision events 
and integrated over the reaction time. For a better resolution of 
the scatter plot we show only about 10$\%$ of the absorption points 
taken randomly. 
Only DCM results are presented, since the UrQMD gives the same 
qualitative predictions. 

In this plot and the following figures we show the elementary hadron 
interactions 
which are responsible for the production of the absorbed 
$\Lambda$-hyperons. This information is important to understand the 
physics of the absorption process. One can see that the contribution of direct 
nucleon-nucleon and pion-nucleon interactions is rather small, since the 
kinetic energy of produced hyperons is quite high in the rest frame 
of the nuclear residues. Only few hyperons emerging from these interactions 
may be captured by spectators. The largest number of absorbed hyperons is 
produced in the secondary interactions of strange particles with nucleons, 
e.g., in strangeness exchange reactions induced by antikaons and in the 
rescattering of fast hyperons on slow spectator nucleons. 

%\vspace{5mm}

\begin{figure}[tbh]
\includegraphics[width=0.8\textwidth]{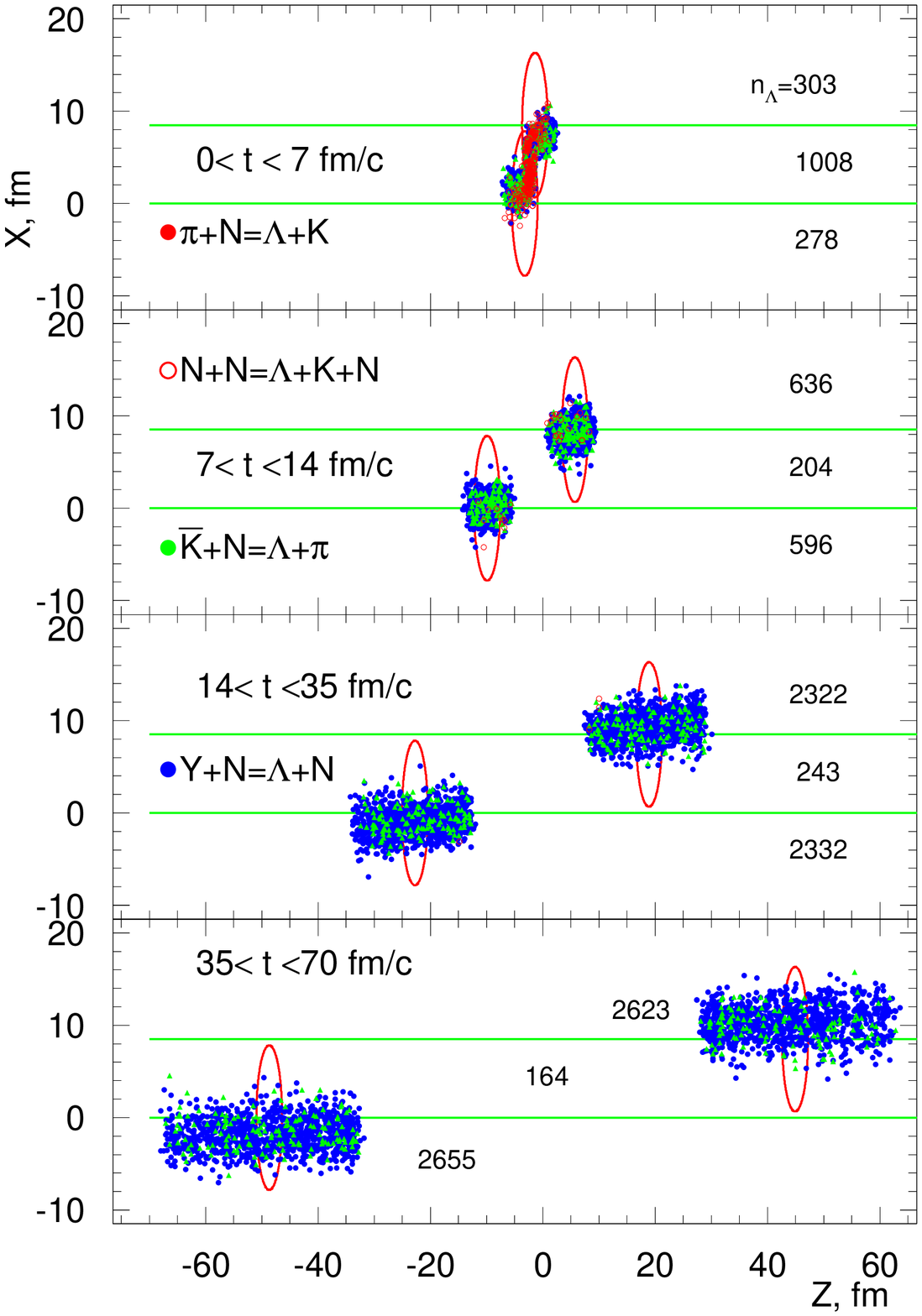}
\caption{\small{
The same DCM calculations with symbol notations as in Fig.~\ref{fig2}, 
but in the X--Z plane. The 
symbols show the coordinates of $\Lambda$ absorption points in projectile and 
target spectators. 
Ellipses show the average positions of projectile and target nuclei during 
time intervals indicated in the figure (from top to bottom) 
in the system of equal velocities. 
Number of hyperons $n_{\Lambda}$ (per 2$\cdot$10$^5$ events) captured in the 
participant and spectator zones during these intervals are noted on the 
right side. 
}}
\label{fig3}
\end{figure}

It is instructive to investigate the absorption rate as a function of 
time to extract the coordinates of absorption points along the beam axis. 
The correlation between the time and X--Z coordinates of the absorption 
points is illustrated in Fig.~\ref{fig3}. 
One can see that some absorption events happen at a very early time 
($\loo$ 10 fm/c), when hyperons produced in the hot participant are captured 
by a few nucleons located in this zone. However, the 
contribution of this mechanism to the total capture yield is rather small. 
Moreover, due to intensive interactions taking place in the participant zone, 
many of the surrounding nucleons will be kicked out at later stages and the 
hyperons can become free again. This is why in the DCM calculations we have 
introduced a multiple check of the absorption criterion at later time steps. 

After the first 10 fm/c, when the 
projectile and target residues are completely separated in coordinate 
space, the capture process is entirely associated with the interactions of 
strange particles in the non-overlapping zones of colliding nuclei. 
Reactions induced by antikaons and hyperons with velocities close to 
projectile and target velocities provide the dominant contribution. 
One can see from Fig.~\ref{fig3} that this process continues for tens of 
fm/c, and it is quite possible to produce a hypernuclear residue even at the 
time of around 50 fm/c. By looking at the X-coordinates of the absorption 
points one can clearly see that the region of the $\Lambda$ absorption moves 
from the overlapping zone at early time to the non-overlapping spectator 
parts for later times, as a result of many secondary interactions. 

\begin{figure}[tbh]
\includegraphics[width=1.0\textwidth]{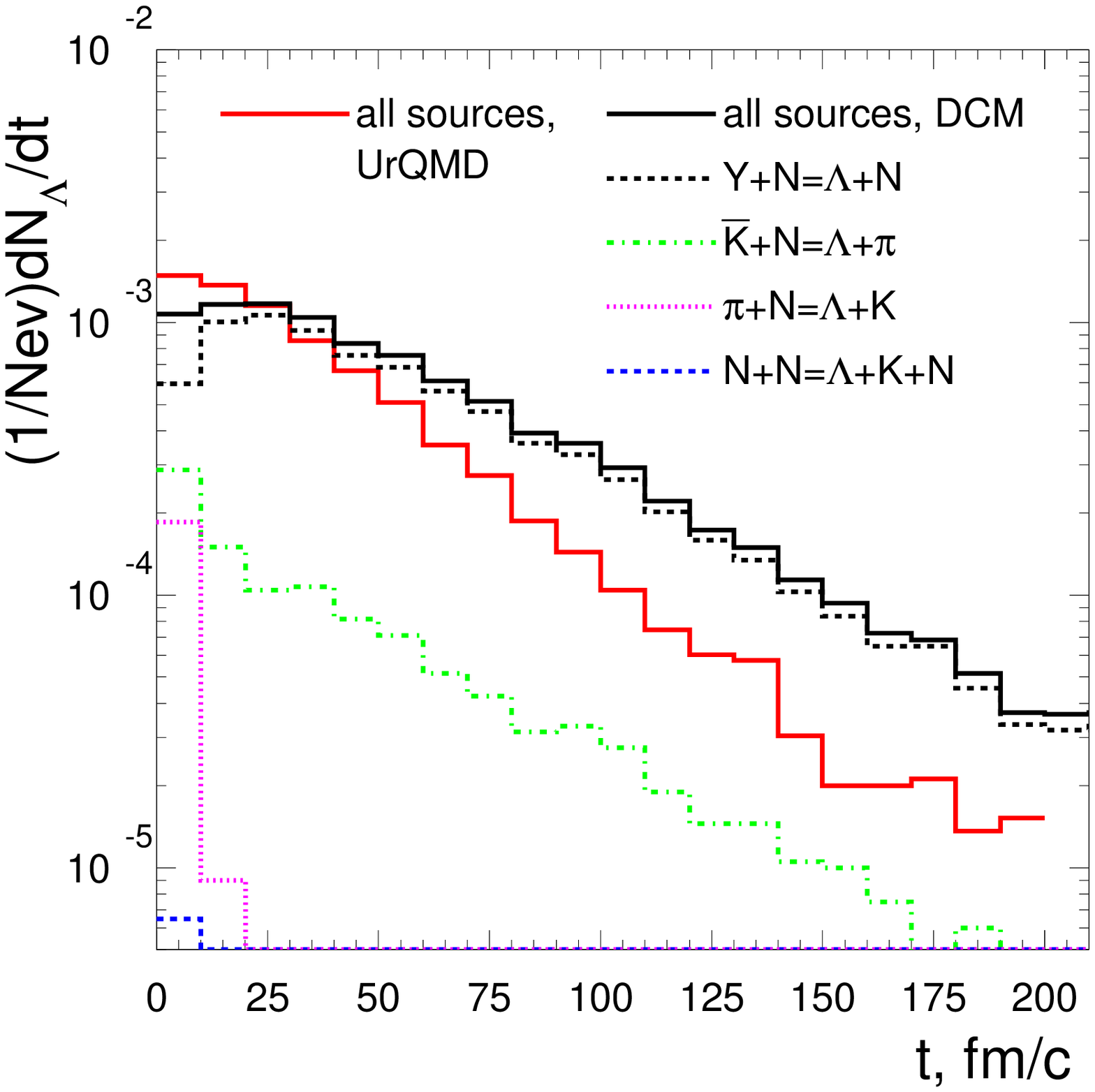}
\caption{\small{
The rate of the $\Lambda$ absorption by spectators (versus time) 
as calculated in DCM and UrQMD model for Au (20 A GeV) + Au reaction. 
Black, green, blue and violet 
lines give the total rate and contributions of specific 
reactions to production of these $\Lambda$ as predicted by DCM. Red line 
is the total rate predicted by UrQMD. 
}}
\label{fig4}
\end{figure}

The space-integrated time evolution of the $\Lambda$ absorption is 
presented in Fig.~\ref{fig4} for the same Au+Au collisions 
at 20 A GeV projectile energy. For convenience, the capture 
rate is normalized to the total number of collision events. This figure 
allows  for a quantitative estimate of the absorption rates associated 
with different interaction channels. From the 
DCM results one can see that about 90$\%$ of $\Lambda$s absorbed in 
spectators come from secondary reactions when previously produced hyperons 
re-scatter on spectator nucleons. 
After the first 
50-70 fm/c the absorption rate becomes low and it drops exponentially, 
signalling that the dynamical process is coming to an end, and the 
equilibrium stage of the spectator's evolution should be considered. 

One can notice a difference between the two considered models: the 
absorption rate decreases faster in the UrQMD model. 
The reason could lie in the treatment of $\Lambda$ absorption at early 
times: In the UrQMD $\Lambda$-hyperons are considered 
as finally absorbed after the capture criterion is first 
satisfied. No further propagation of these hyperons is simulated. 
As we have mentioned, in the DCM we take into account a time evolution 
of the nucleon density in the vicinity of the captured hyperon. Some 
nucleons can leave this neighbourhood at a later time, leading to a decrease 
of the local density so that the hyperon may escape. This happens mostly 
at the early stage ($\loo 10$ fm/c). 
On the other hand, at later times some of these hyperons may be captured 
again in the spectator zone, and this effect will increase the 
absorption rate. Therefore, the absorption process is longer in the DCM 
as compared with the UrQMD. However, in the end these two dynamical 
scenarios of hyperon absorption give very similar integrated yields.

%\vspace{5mm}

\section{Characteristics of spectator hyper-matter.}

\begin{figure}[tbh]
\includegraphics[width=1.0\textwidth]{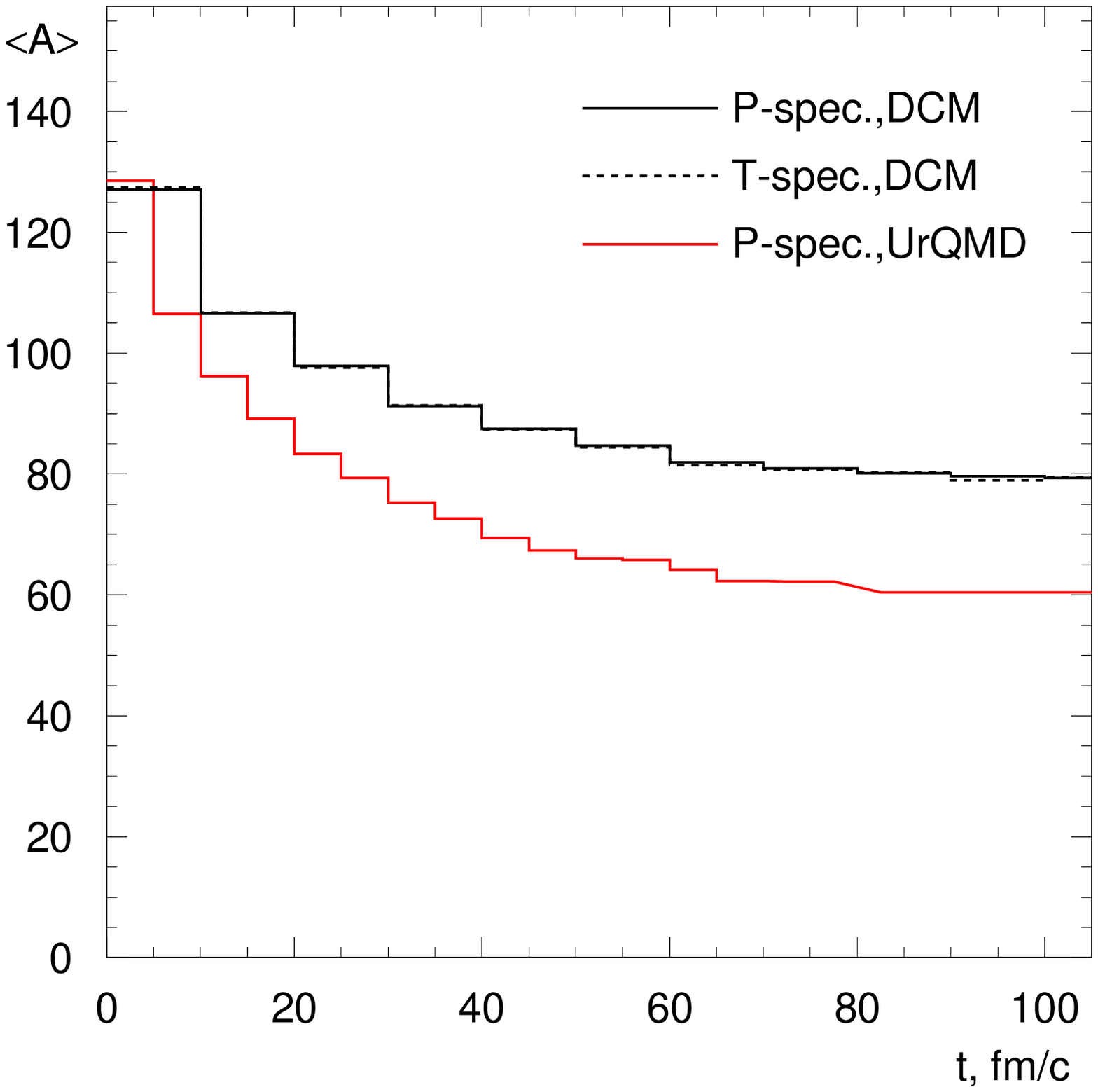}
\caption{\small{
Time evolution of the mean number of nucleons in the spectator 
(P -- projectile, T -- target) residual 
nuclei after $\Lambda$ absorption as predicted by the DCM and UrQMD 
calculations. The reaction is as in Fig.~\ref{fig2}. 
}}
\label{fig5}
\end{figure}

As one can see in Fig.~\ref{fig4}, 80--90$\%$ of the $\Lambda$s are 
absorbed in 
spectators by the time of 80 fm/c. At this stage all 
secondary interactions are practically over and we can determine 
the parameters of the spectator residues. 
Figure~\ref{fig5} shows the time evolution of the average mass number 
of spectators as predicted by DCM and UrQMD calculations. 
The decrease of masses with time is obviously caused by secondary interactions 
of fast particles with spectator nucleons. In the UrQMD a nucleon is 
defined as spectator if it has not participated in elementary interactions. 
On the other hand, the DCM definition of spectators includes nucleons which 
underwent interactions and have been recaptured by the nuclear potential. 
By this reason the DCM spectator residues are bigger. 
It is seen that the total number of nucleons in the spectators is nearly 
saturated after 70 fm/c. During the dynamical stage the high energy neutrons 
and protons interact similarly, therefore, the average neutron-to-proton 
ratio in the spectators is almost the same as in colliding nuclei. 

These spectator residues acquire some excitation energy, 
which can be calculated within both models too. However, it is well 
known from experiments with conventional nuclear matter that in 
relativistic collisions the average temperature of such residues is saturated 
at the level of 5--6 MeV, despite of large event-by-event fluctuations 
of the excitation energy \cite{smm,pochodzalla}. We expect similar 
average temperatures in 
hyper-residues too, especially when the number of captured hyperons is 
much smaller than the number of nucleons.

\begin{figure}[tbh]
\includegraphics[width=1.0\textwidth]{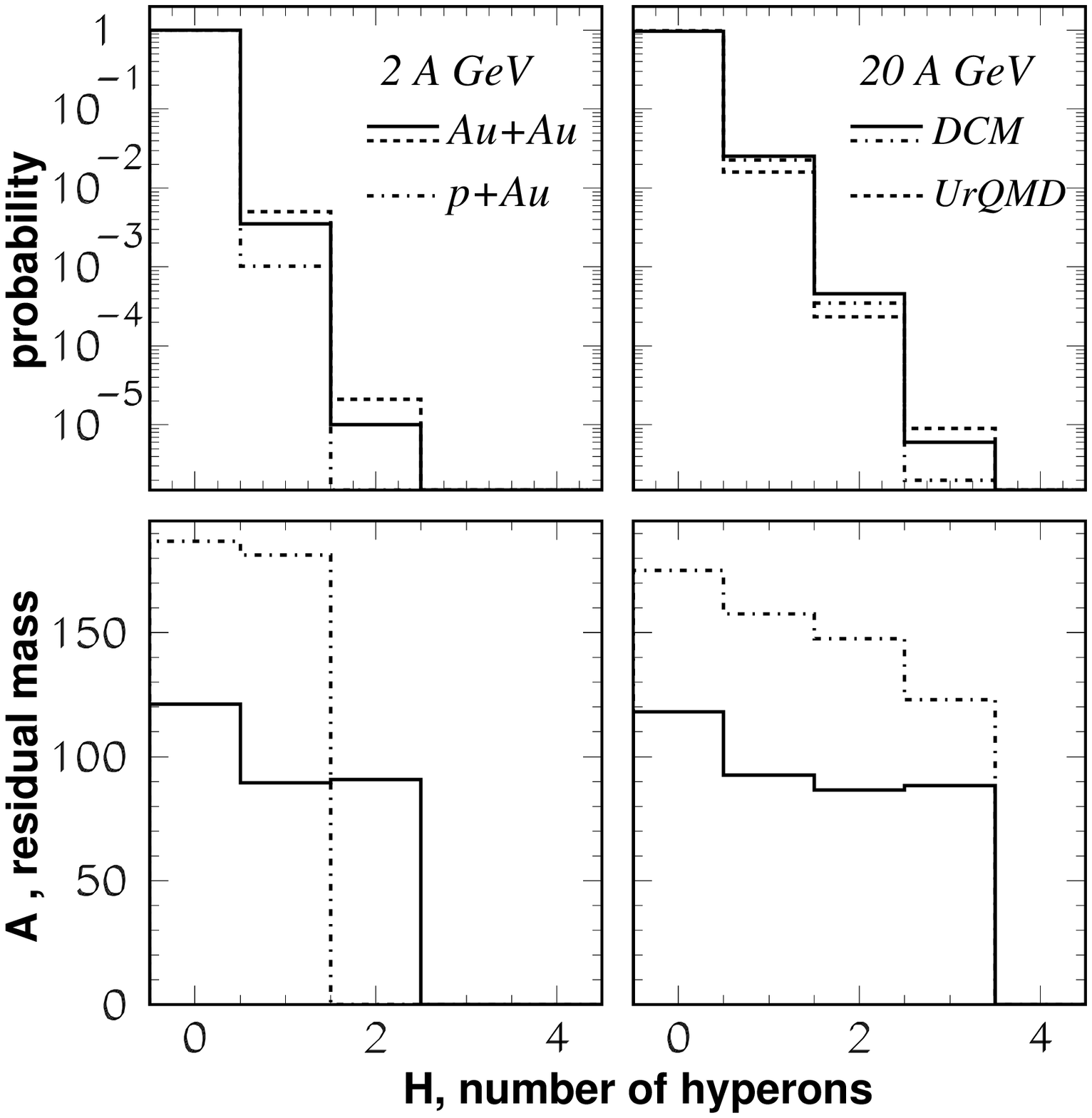}
\caption{\small{
Probability for formation of conventional and strange spectator residuals 
(top panels), and their mean mass numbers (bottom panels)
versus the number of captured $\Lambda$ hyperons (H), calculated with DCM 
and UrQMD model for p + Au and  Au + Au collisions with energy of 
2 GeV per nucleon (left panels), and 20 GeV per nucleon (right panels). 
The reactions and energies are noted in the figure by different 
histograms. 
}}
\label{fig6}
\end{figure}

One important practical result of this study is the estimated probability 
for producing spectator residues with different numbers of captured 
$\Lambda$s. These probabilities are shown in Fig.~\ref{fig6} for residues 
containing up to 3 $\Lambda$ hyperons, together with their mean masses, 
for Au + Au and p + Au collisions with energies of 2 and 20 A GeV. 
The chosen beam energies correspond to the 
present GSI experiments, and the future FAIR experiments. The calculations 
were performed with minimum bias, i.e., integrated over 
all impact parameters. For each reaction 10$^6$ Monte-Carlo events were 
generated.  
According to the DCM calculation, at the highest energy the probability for 
capturing one hyperon in the spectator amounts to a few percents, 
and two hyperons may be captured with a significant probability of about 
$\sim 5\cdot 10^{-4}$. Furthermore, even 3 $\Lambda$-hyperons may be captured 
by spectators 
with a low but measurable probability $\sim 10^{-5}$. The absorption of a 
higher number of hyperons is also feasible. This new mechanism opens a 
unique opportunity to produce and study multi-strange systems, which are not 
conceivable in other nuclear reactions. Since the predicted masses of the 
hyper-spectators are quite large, one can even speak about the formation of 
excited hypermatter. We have found that variation of the hyperon potential 
within a reasonable range does only weakly influence the capture rates 
at the highest energy, because these hyperons undergo multiple 
rescattering and de-accelerate considerably in the spectator matter. 

Compared to the DCM results, the probabilities for producing hyper-spectators 
predicted by the UrQMD model can differ by a factor 2. 
This difference shows the uncertainty in the calculations based on the 
best theoretical models available at present. Still the qualitative 
agreement between the two transport models gives us a confidence that 
this new method can be successfully used for producing new hypernuclei. 

From calculations with the UrQMD model we have also estimated the probability 
of a $\Xi$ being absorbed in the spectator fragment at $E_{lab}=20$ A GeV 
to be on the order of $10^{-5}$ per event. Such an absorption would 
eventually lead 
to the formation of $\Xi$--hypernuclei in future experiments. This 
opens the interesting possibility to study a new multi--strange system which 
is even less understood as the conventional $\Lambda$--hypernuclei.

\begin{figure}[tbh]
\includegraphics[width=1.0\textwidth]{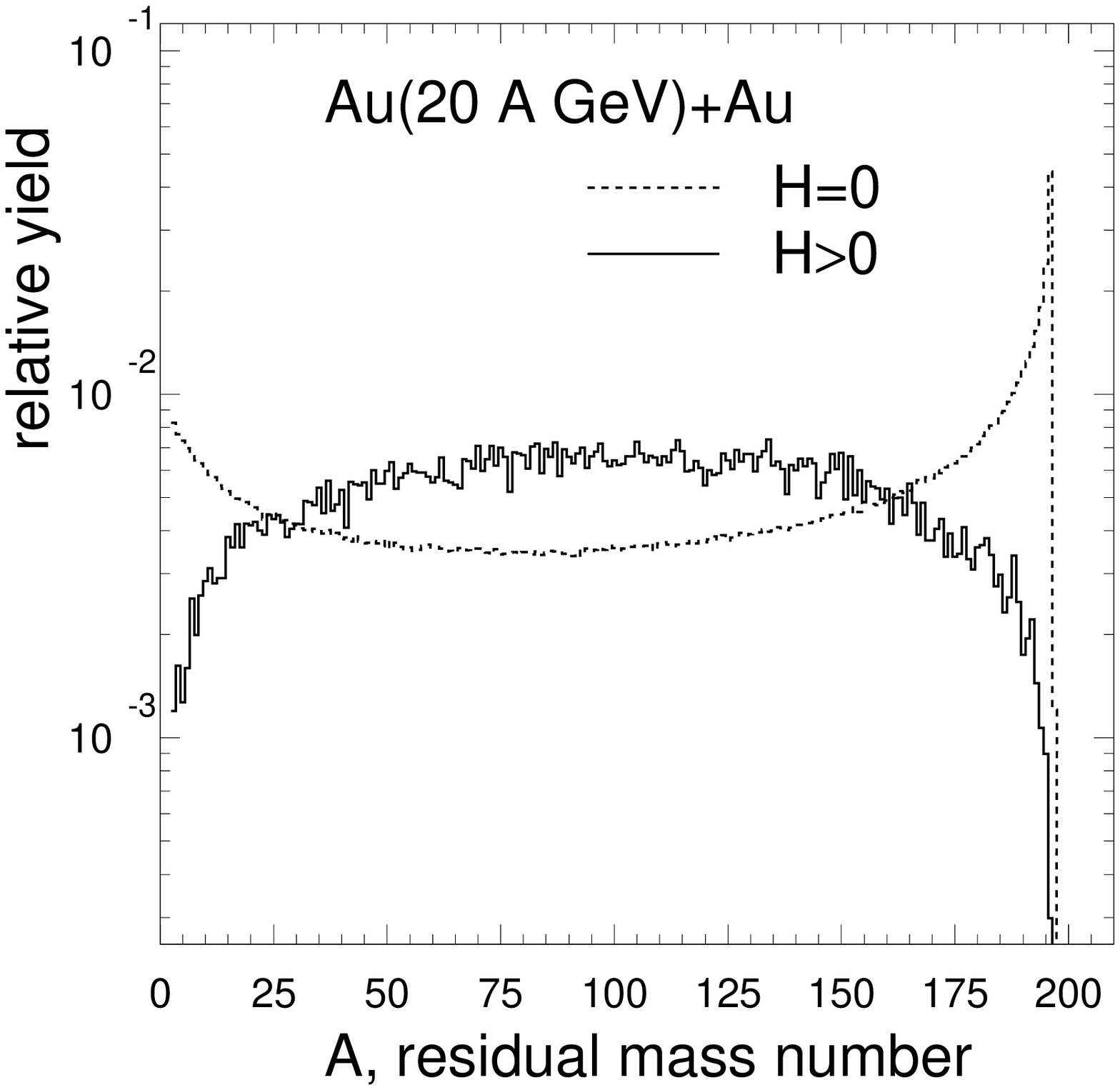}
\caption{\small{
Mass distribution of conventional (H=0) and hypernuclear (H$>$0) spectators 
after DCM calculations for Au + Au collisions with projectile energy 20 
GeV per nucleon. 
}}
\label{fig7}
\end{figure}

For completeness, in Fig.~\ref{fig7} we show the DCM predictions for 
the mass distributions of residues (representing both conventional matter 
and hyper-matter), integrated over all impact parameters. One can notice 
a significant difference in the shape: for conventional residues 
the distribution is very broad, and it has a sharp peak around the 
projectile mass, corresponding to very peripheral collisions. These heavy 
residuals are weakly excited since only a few nucleons participated in the 
reaction. On the contrary, when hyperons are absorbed many nucleons are 
involved in secondary interactions, and some of them leave the spectators. 
As a result, it is very unlikely to produce hyper-residues with masses 
around the initial mass in heavy-ion collisions. However, the distribution 
remains very broad and practically all intermediate-mass residues can be 
produced with considerable probability. 

%\vspace{5mm}

We should note that the formation of hyper-residues was earlier discussed 
in Ref.~\cite{cassing} for the reactions initiated by protons. 
As one can see from Fig.~\ref{fig6} the probabilities for production of 
multi-strange hyper-residues in such reactions are lower 
than in nucleus-nucleus collisions. This is easy to understand, since 
more hyperons in the spectator region can be produced when more particles 
participate in the interactions. However, the channels with absorption 
of one $\Lambda$, and even with two $\Lambda$s at the highest energy 
can be still very probable in p + Au collisions. In addition, the masses of 
hyper-residues will be larger, since in this case less target nucleons are 
involved in the reaction. The main problem expected for proton beams is that 
experimentally it is very difficult to identify decay products of a slow 
hyper-nucleus in a background of free $\Lambda$ decays. Relativistic 
heavy-ion collisions have essential advantages: Because of the Lorentz 
%$\gamma$ 
factor their lifetime becomes longer and projectile hyper-fragments can 
travel a longer distance. This makes possible to use sophisticated vertex 
detectors and fragment separation technique for their identification. 

%\vspace{5mm}

\section{Rapidity distributions of hyper-residues and free $\Lambda$-hyperons}

\begin{figure}[tbh]
\includegraphics[width=1.0\textwidth]{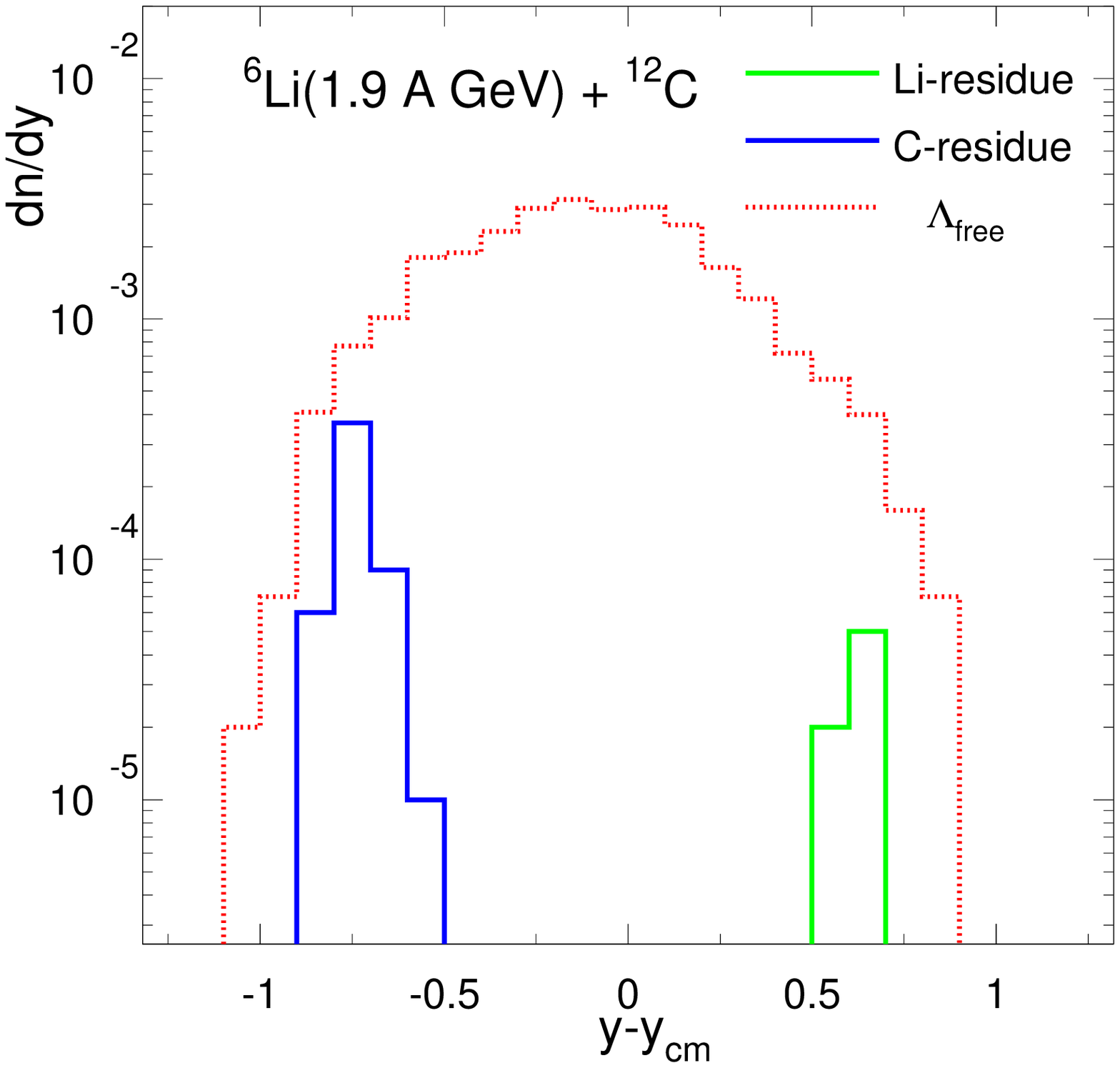}
\caption{\small{
DCM calculations of rapidity distribution for free $\Lambda$ and for 
hyper-residues with one $\Lambda$ coming from target (C) and projectile (Li) 
spectators in reaction $^{6}$Li (1.9 A GeV) + $^{12}C$. 
}}
\label{fig8}
\end{figure}

For the experimental identification of hypernuclei it is of crucial 
importance to know 
the background associated with free hyperons produced in a reaction. 
For example, let us consider the reaction $^{6}$Li + $^{12}$C at a beam 
energy of 1.9 GeV per nucleon, 
which was employed on the first stage of the HypHi 
experiment at GSI \cite{hyphi}. Figure~\ref{fig8} shows the rapidity 
distributions of the free and captured $\Lambda$s produced in this 
reaction. One can see that for these light nuclei the 
yield of free $\Lambda$s in the spectator region is 1--2 orders of 
magnitude larger than the yield of hypernuclei originating from the 
projectile spectator. 
For this reason, the fraction of pions coming from the decay of hypernuclei 
is very small and careful correlation measurements are required to identify 
these hypernuclei. It is interesting 
that this fraction is higher for the larger target spectator, however, 
in the experiment only projectile fragments can be detected. 

\begin{figure}[tbh]
\includegraphics[width=1.0\textwidth]{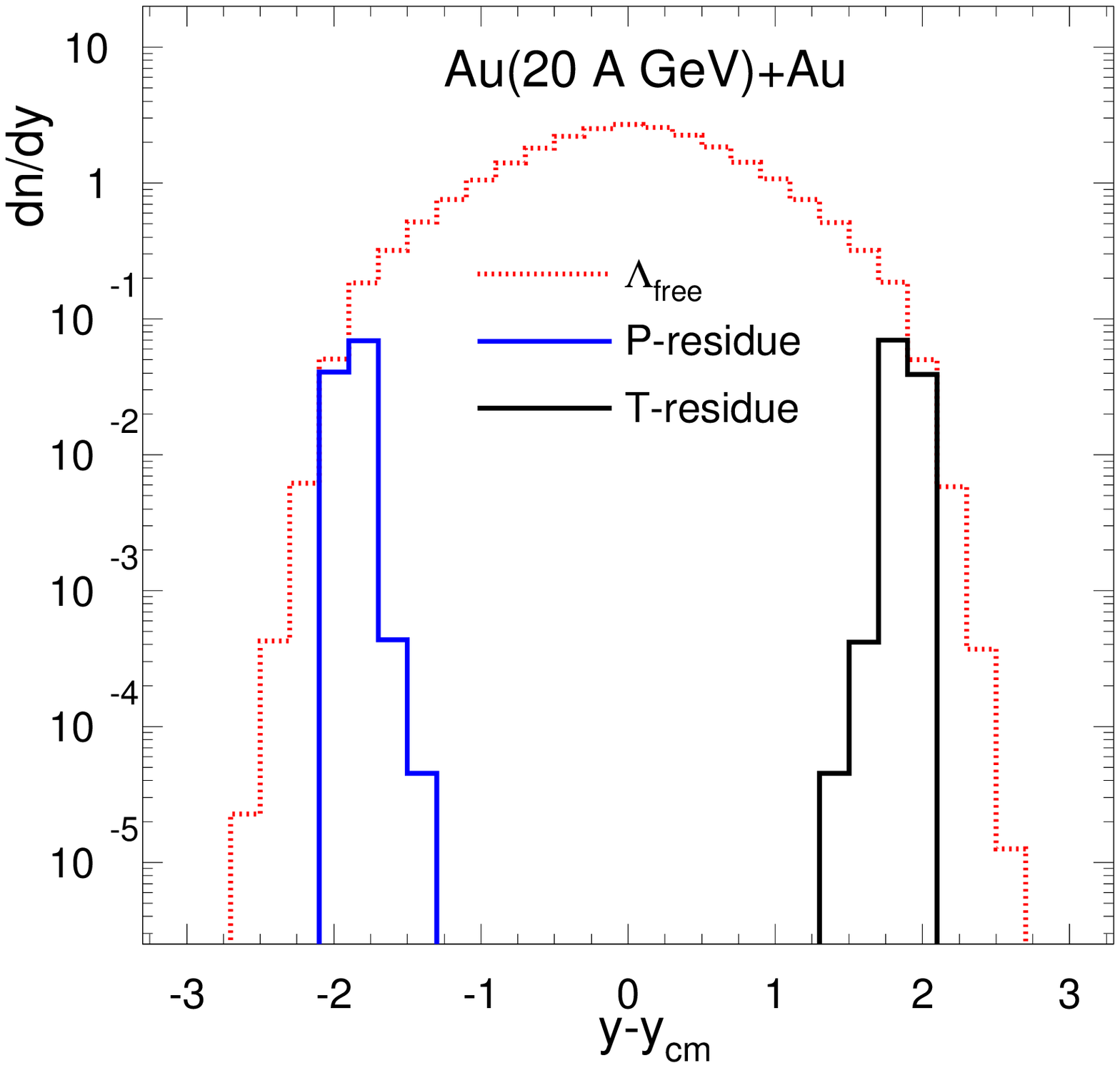}
\caption{\small{
DCM calculations of rapidity distribution for free $\Lambda$ and for 
hyper-residues coming from target 
(T) and projectile (P) spectators in the reaction 
$^{197}$Au (20 A GeV) + $^{197}$Au. 
}}
\label{fig9}
\end{figure}

A higher ratio of spectator hypernuclei to free $\Lambda$s 
is obtained in reactions involving heavy nuclei. Figure~\ref{fig9} 
presents the result of the DCM calculation for Au + Au collisions at 
a beam energy of 20 A GeV, as planned in future FAIR experiments. 
One can see clearly that the excess of free $\Lambda$s over hypernuclei in 
spectator kinematic zones is minimal in this case, i.e., less than a 
factor $2 \div 3$. This significant improvement has a simple explanation: 
More secondary hyperons interact with spectator 
nucleons at a later time of the reaction ($t > 20$ fm/c). These hyperons 
have velocities close to the velocities of spectators, therefore, it is 
more probable for them to have other spectator nucleons nearby. 

%\vspace{5mm}

\section{Momentum distribution of absorbed hyperons.}

\begin{figure}[tbh]
\includegraphics[width=1.0\textwidth]{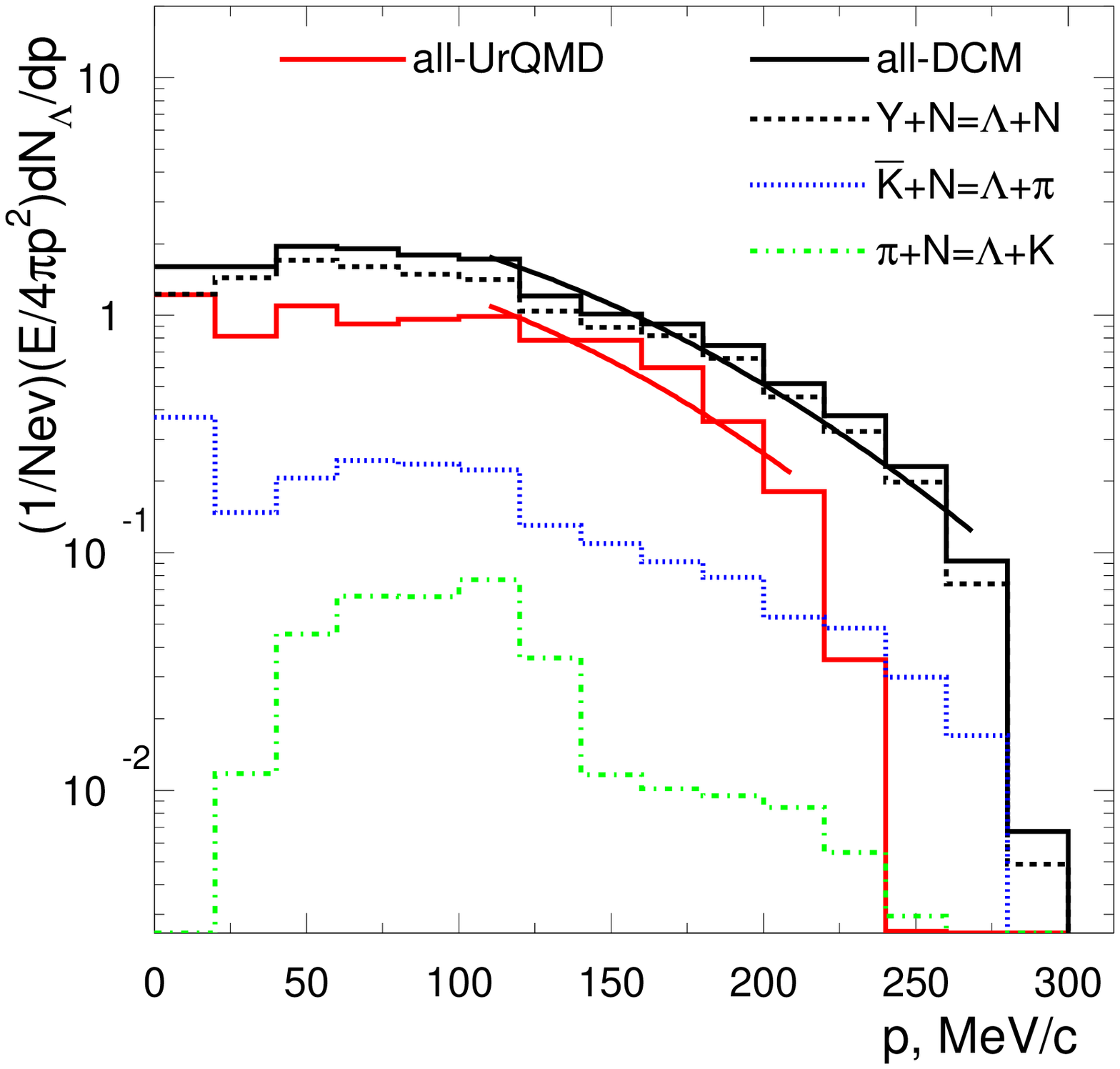}
\caption{\small{
Momentum distribution of $\Lambda$ hyperons captured in 
the spectators. Notations are the same as in Fig.~\ref{fig4}. 
Solid curves show the Boltzmann fits of the distributions with 
temperature $\sim 10$ MeV. 
}}
\label{fig10}
\end{figure}

As follows from the potential criterion (see Section {\bf II}) the absorbed 
$\Lambda$ hyperons have small momenta in the rest frame of the spectators. 
Direct DCM and UrQMD calculations of the invariant momentum distribution 
of the absorbed $\Lambda$s for Au (20 A GeV) + Au collisions 
are shown in Fig.~\ref{fig10}.
One can see that the distribution is rather narrow with a smooth 
cut-off for hyperon momenta greater than 250$\div$300 MeV/c. 
However, this distribution is not step-like, and we can even 
approximate the slope of the spectrum at $P > 100$ MeV/c by a 
Boltzmann distribution with apparent temperature 
$T \approx 10$ MeV. This low temperature indicates that these 
hyperons are rather slow in the considered reference frame, and this is 
a typical temperature for particles coming from the spectator matter as 
was demonstrated long ago by the analysis of ALADIN data \cite{aladin97}. 

Sometimes the coalescence criterion for cluster formation is used, 
when a hyperon is considered to be absorbed by a nuclear cluster if they 
are close in momentum and coordinate space. Our results shown in 
Fig.~\ref{fig10} justify using the coalescence criterion in a simple 
consideration, when secondary interactions in spectators are neglected. 
Recently the GiBUU model \cite{gibuu} was generalized to describe 
hypernuclei production in similar relativistic nucleus-nucleus 
collisions. They have calculated excited residues after the dynamical 
stage. Then the SMM \cite{smm} was applied to model the de-excitation 
and multifragmentation of these residues into conventional fragments. 
Afterwards, a coalescence prescription to form hyperfragments from these 
conventional fragments and free $\Lambda$s was applied. 
An important feature of our approach is the assumption that the 
absorption takes place fastly (during first few tens of 
fm/c) in excited residues before their break-up. The de-excitation 
itself takes a considerable time, of about 10$^{2}$--10$^{4}$ fm/c 
depending on excitation energy. Therefore, the final yields of 
specific hypernuclei in our approach will be different from the 
approach of Ref.~\cite{gibuu}.

Nevertheless, the coalescence model can be successfully used to take into 
account the final state interaction between free nucleons leading to the 
formation of lightest clusters, as was shown already in Ref.~\cite{toneev83}. 
We expect that this mechanism may work in the case of light spectators 
and free hyperons too. 
Similar results concerning the production of light hyper-fragments can be 
obtained within the thermal models 
\cite{BraunMunzinger:1994iq,Steinheimer:2008hr,stoecker}. 
A connection between parameters of coalescence and thermal models describing 
the production of lightest clusters at high excitation energies is under 
discussion for a long time, see, e.g., Refs.~\cite{mekjian,neubert}. 
In principle, the coalescence prescription can be applied for free nucleons 
and hyperons to investigate the formation of both, conventional light nuclei 
(such as $d$, $t$, $He$), and light hyper-nuclear fragments produced in the 
spectator region. 
However, our preliminary calculations for Au +Au collisions have shown that 
a probability for the production of hyper-clusters by this mechanism is by 
several orders of magnitude smaller than the absorption of hyperons in the 
spectators, as considered in this paper. On the other hand, we have found that 
the coalescence mechanism is more efficient for the formation of 
hyper-clusters in the midrapidity zone \cite{coales}.

%\vspace{5mm}

\section{Disintegration of hot spectator matter into hyper-fragments.}

In the final stage of the reaction the excited hyper-residues undergo 
de-excitation. At low excitation energy, this should be an evaporation 
(and, may be, fission) process, similar to what is well known for the 
conventional nuclei. However, as we see from our calculations and also from 
analyses of experimental data \cite{smm,aladin97,EOS}, the thermalized 
residual nuclei have rather high excitation energies in the nuclear scale, 
and, consequently, they should undergo multifragmentation with a 
characteristic time of about 100 fm/c.  
The generalization of the Statistical Multifragmentation Model 
(SMM) \cite{smm} into the strangeness sector by including $\Lambda$-hyperons 
has been done in Ref.~\cite{bot-poch}. 
It was demonstrated that the fragment mass distributions are
quite different for fragments with different strangeness content. This means
that the multifragmentation of excited hypernuclear systems
proceeds in a different way as compared with conventional
nuclei. The reason is the additional binding energy of hyperons in nuclear 
matter. It was also shown  
that the yields of fragments with two $\Lambda$s
depend essentially on the mass formulae (i.e., on details of $\Lambda N$ and 
$\Lambda \Lambda$-interactions) used for the calculations 
\cite{bot-poch,samanta}. Therefore, an analysis of double hypernuclei can help 
to improve these mass formulae and reveal information about 
the hyperon-hyperon interaction. In Ref.~\cite{lorente} the 
decay of light excited hyper-systems was considered within the framework 
of the Fermi break-up model. It was also concluded that the 
production rate of single and double hyper-nuclei is directly related to 
their binding energy.  

We would like to mention another canonical statistical model 
\cite{dasgupta}, which was developed in line with \cite{bot-poch}. 
It gives similar results 
concerning hypernuclei produced after the decay of excited hyper-systems. 
A detailed analysis of the de-excitation process taking into account the 
primary dynamical stage of the reaction will be presented in a 
subsequent publication. 

%\vspace{5mm}

\section{Conclusion}

Within the DCM and UrQMD models we have investigated the production of 
hyperons in peripheral relativistic heavy ion collisions and their capture 
by the attractive potential of spectator residues. 
Contrary to the coalescence mechanism, 
which may be responsible for the formation of light hyperfragments, 
this capture process can also lead to the production of heavy hypernuclei. 
Yields of such hypernuclei, are quite significant, and a very broad mass 
distribution of hypernuclear spectators can be obtained. 
An important advantage of this method over the traditional reactions 
induced by kaons is that it provides a natural way to produce multi-hyperon 
systems. As our calculations show, the relative probabilities to produce 
residual nuclei with 2$\Lambda$ and 3$\Lambda$ in Au + Au reaction at 
20 A GeV are about 5$\cdot$10$^{-4}$ and 10$^{-5}$, respectively. These 
probabilities should be sufficient for systematic studies of such systems 
in the future FAIR and NICA experiments. 

Conventional spectator matter after relativistic nuclear 
collisions was intensively investigated during the last 20 years 
\cite{smm,pochodzalla}. 
Similarly, we expect the production of excited hypermatter 
at a baryon density $\rho \sim 0.1-0.8 \rho_0$ and temperatures 
around $T \sim 3-7$ MeV, which are typical for the coexistence region of 
the nuclear liquid-gas phase transition. This opens the possibility to study 
this phase transition in nuclear matter with a strangeness admixture, 
that is important for the physics of neutron stars. An analysis of the 
disintegration of this matter into cold non-strange fragments and 
hyper-fragments will 
reveal information about the properties of hypernuclei, their binding 
energies, and, finally, $YN$ and $YY$ interactions. 
Hypernuclei of all sorts are expected to be 
produced in his way, including nuclei around drip lines, which is not 
possible with traditional methods. 

It would be very helpful to perform further detailed theoretical 
calculations, which include all stages of the reaction: 
1) production of hyperon-rich spectator nuclear systems, 2) formation of 
hyper-fragments via decay of these systems, 3) decay of the hypernuclei 
via weak interaction with production of non-strange particles in the final 
state. The DCM and UrQMD approaches can be used for the description of the 
dynamical stage and they can be followed by statistical models, e.g., as 
the SMM. 
In this case all products of the reaction can be determined on an 
event-by-event basis. Advanced correlation measurements, which are necessary 
to identify hypernuclei, can be tested with these models too. 

The detection of spectator hyper-nuclei may suffer from a background 
problem, which may require sophisticated experimental methods to solve. 
However, it has already been demonstrated experimentally that in peripheral 
collisions of light projectiles and targets the cross section for 
the capture of produced $\Lambda$-hyperons by projectile residues may 
be of the order of a few microbarns \cite{Avr88}, and this cross 
section allows for an experimental identification of the produced 
hypernuclei \cite{hyphi}. As we have demonstrated, involving heavy 
projectiles and targets can indeed improve the signal-to-background 
ratio significantly. 
We believe that hypernuclear physics will benefit strongly from 
exploring new production mechanisms and the novel detection technique 
associated with spectator fragmentation reactions. 
 
\begin{acknowledgments}
This work was supported by BMBF, HGS-HIRe and the Hessian LOEWE initiative 
through the Helmholtz International center for FAIR (HIC for FAIR). 
We acknowledge also partial support from the DFG grant 436 RUS 113/957/0-1, 
as well as the grants RFBR 09-02-91331 and NSH-7235.2010.2 (Russia). 
K.K.G. thanks FIAS for the hospitality. 
\end{acknowledgments}

\end{document}